# Interface study of FeMgOFe magnetic tunnel junctions using 3D Atom Probe

B.Mazumder, M.Gruber, A.Vella, F.Vurpillot and B. Deconihout
*Groupe de Physique des Materiaux, University of Rouen,France*

**Abstract:**

A detailed interface study was conducted on a Fe/MgO/Fe system using laser assisted 3D atom probe. It exhibits an additional oxide formation at the second interface of the multilayer structure independent of laser wavelength, laser fluence and the thickness of the tunnel barrier. We have shown with the help of simulation that this phenomena is caused by the field evaporation of two layers having two different evaporation

## 1. Introduction

Magnetic tunnel junctions (MTJs) consisting of two ferromagnetic metal layers separated by a thin insulating oxide barrier have attracted an increasing interest in the last decade due to their potential applications in magneto electronic devices [1-3]. When a bias voltage is applied between the two metal electrodes, electrons can tunnel through the thin barrier, inducing a tunneling current, which is dependent on the relative orientation of magnetization of the two ferromagnetic layers. This phenomenon is called tunneling magneto resistance (TMR) effect, which is one of the most important effects in the field of spintronics. MTJ's consisting of amorphous aluminium oxide ($Al_2O_3$) tunnel barrier and ferromagnetic electrodes with high spin polarization [4-7] show relatively low (TMR) ratio, of up to about 70% at room temperature , which limits the implementation of such structures in spintronic devices. However, theoretical calculations[8, 9] show high magneto resistance of up to 1000% for Fe/MgO/Fe MJT's with abrupt junctions. Experimentally, MgO-based MTJs fabricated by various groups have shown TMR ratios as high as over 200% at room temperature [10-12]. This discrepancy between theoretical and experimental values of TMR is probably due to the interface properties of the heterostructure.

Hence, the investigation of growth and interface properties of Fe/MgO/Fe becomes crucial for improving the performance of MTJs. Various groups have simulated the interface between Fe/MgO using first principle calculations [13, 14]. Since these are all based on assumptions and ideal sample conditions, these studies are not in agreement with results from the actual interface study obtained using different methods like RHEED [15] and XPS [16] which give two dimensional information for the interface area.

Among the modern relevant approaches in nano science, Atom Probe Tomography (APT) provides key information in the area of nanotechnology and nanoscience due to its uniqueness in three dimensional atomic scale imaging for nanodimensional materials. This technique has enormous potential for giving information in atomistic scale with a sub nanometre spatial resolution. Various applications of this technique have been demonstrated in literature [17]. Earlier, the method was unique to metals and semiconductors but now is successfully applied to insulators with the advancement of pulsed laser enhanced evaporation [18, 19]. It is now possible to have a good 3D atom probe reconstruction for high band gap materials like $Al_2O_3$ [20] and $HfO_2$ [21] . High band gap materials like MgO, having band gap of 7.16 eV have been sparsely studied by APT to date [22, 23].

In this work, we have studied the interface between metallic layer Fe and insulating layer MgO with respect to different laser wavelengths. The inter diffusivity mechanism at the interface is discussed and well established by theoretical model.

## 2. Experimental methods
### 2.1 Deposition of MgO layer

The multi layered structures of Fe/MgO/Fe are prepared by means of argon plasma sputter deposition on a prepatterned substrate consisting of an assembly of flat-topped Si (100) pillars (5×5×100) $\mu m^3$. Prior to deposition, the substrate is cleaned for 10-15 min in a standard solution of $H_2O_2+H_2SO_4$ and after 10min in another solution of 10% HF to remove residual contamination and silicon oxide. It is then thoroughly rinsed with distilled water and dried in stream of dry nitrogen. After, the layers to be investigated are deposited within the sputter deposition chamber. The schematic of the sputter deposition setup used is depicted in Fig.1[24]. It is very important to clean the substrate by ion beam prior to deposition to achieve good mechanical stability of the specimen [24]. The base pressure of the deposition chamber before deposition is $10^{-5}$ Pa. Substrate, as well as each target cleaning is done with 500 V $Ar^+$ ions and a beam current of 15 mA for 20 s (step 1 in Fig. 1). Ar pressure of the system during the sputtering is 2 x $10^{-2}$ Pa. The MgO layer is deposited using an MgO target with no further additional oxidation and without any oxygen floating during deposition. The sputtering rate for Fe and MgO are 2.26nm/min and 0.3nm /min respectively. The first and third layers sputtered are Iron (Fe), in a thickness of 30 nm and in between, the insulating barrier of MgO is deposited with a thickness of 4nm on one sample and 32 nm on the second one.

## 2.2 Tip preparation and measurement

The silicon pillars are removed from the wafer and mounted on a fine tungsten needle with conductive commercial adhesive and a fine micro- manipulator under the optical microscope. They are then subjected to the Focussed Ion Beam (FIB) and shaped within an optimised procedure into tips suitable for APT analysis. To avoid damages in the region of interest and reduce Ga implantation the multilayer is capped with a Cr layer of 500-600 nm before exposure to Ga beam. Figure 2a and 2b shows the scanning electron microscopy (SEM) micrograph of the specimen with different thickness of MgO layer prior to analysis. These pictures give evidence of a geometrically appropriate tip with dark and bright contrast indicating the different electron density between the layers. Clearly, dark Fe layers and white MgO layer are seen in the image. The specimen of Fig 2a has a two fold tunnel barrier separated by a gold layer of 10nm. However, the exact thickness of the layer can not be estimated. These multi layers are analyzed by Laser Assisted Wide Angle Tomographic Atom Probe (LAWATAP) in an ultrahigh vacuum chamber at a pressure of $10^{-9}$-$10^{-10}$ Pa. Details of the instrument are given in literature [25]. The femtosecond laser pulse system used was an amplified ytterbium-doped laser (AMPLITUDE SYSTEM HR-pulse) with a pulse length of 350 fs and the repetition rate of 100 KHz. The laser system is equipped with three different wavelength sources, IR (1030nm), Green (515nm) and UV (343nm). For our work 515 nm and 342 nm are used.

## 3. Results and discussion
### 3.1. Analysis with UV laser ($\lambda_1$=343 nm)

The first analysis is performed with the sample containing 4 nm thick MgO barrier. Prior to analysis, the specimen was cooled down to 80 K. Analyses were performed using laser fluence of 4 J/m$^2$ focused at the apex of the specimen and fixed evaporation rate (0.005-0.003 atoms/pulse) throughout the experiment. The mass spectrum of thin MgO layer analysed within UV range is shown in Fig 3a. The mass spectrum reveals the expected peaks for single and double charged Mg at ~24 amu and ~ 12 amu, the peaks of single and double charged Fe at 56 amu and 28 amu, $O_2$ at 32 amu and O at 16 amu. A rigorous scrutiny reveals the presence of molecular ions of Mg-O bonds in the analysis. In addition, the mass spectrum in the proximity of the tunnel barrier shows several metallic oxide peaks, such as $FeO^+$ and $Fe_2O$. Also, we observe hydrogen and oxygen peaks showing a slight contamination of the sample from the residual gases $H_2$, $H_2O$ adsorbed from the vacuum preparation or analysis chambers. Even with the laser pulsing, the total dissociation of the molecular ions is not achieved.

The analysis of the specimen with a comparatively thicker layer of MgO (~32nm) was performed with the same protocol as described above. Analyses were performed using laser energy of 10.4 $J/m^2$ with evaporation rate of 0.005-0.003 atoms/pulse. The analysis was successful until the evaporation reaches the second insulator-metal (MgO/Fe) interface. At this interface, the tip ruptures due to the enormous voltage increase. In order to keep the voltage less increasing the laser fluence was increased from 7.5 $J/m^2$ up to 9 $J/m^2$ in the region of the second interface and the entire layer was analysed. Even at 20K where the oxide is more fragile, the analysis was possible by changing the laser energy at the interface. The mass spectra of thick layer of MgO analysis is shown in the Fig 3b. We see the same traces of the major elements and the oxide molecular ions as well. A typical 3D spatial reconstruction of the atom distribution of the thin layer and thick layer of MgO are shown in Fig 4a and 4b respectively. The MgO layer is clearly seen, in agreement with the multilayered structure observed by SEM. But in SEM one can not have the exact measurement of the thickness of the layers whereas in Atom Probe the correct thickness can be measured. The thicknesses of the MgO layers of 4nm and 32nm are well reconstructed. The reconstruction also allows visualization of the interfaces between the different layers. Noticeable things at the interface for both cases are that the lower interface is very much oxidized and diffused and there is an existence of a depleted zone just after the oxide layer.

To study the interfaces the concentration profile was calculated on a parallelepiped volume of $7\times7\times40$ $nm^3$ and $8\times8\times50$ $nm^3$ respectively indicated by red lines in Fig 4a, b. The concentration profiles of elements for the two different layers are shown in Fig 5a and Fig 5b. We have to use a volume for the concentration profile smaller than the total volume because due to the wide angle analysis. The interface is not perpendicular to the tip axis on the all volume. In the first sample (4nm) a high purity ($\approx$ 90%) of Fe concentration is found in the first layer. The Mg and O concentrations show maxima of 44% and 42% respectively. The ratio of Mg and O of 1.04 $\pm$ 0.15 is detected in the center of the oxide barrier shown in Fig 5a.

The O peak is extended to the right of Mg peak suggesting the oxidation of lower ferromagnetic electrode [20, 25]. The presence of the molecular species of FeO and $Fe_2O$ is higher in the lower electrode. There is also a small Fe peak just after the MgO layer but not coinciding with the O peak. The profile for thick layer shows almost the same behaviour as that of the thin layer. In this case, the content of Mg is 48 % in the center of the layer, the Mg content is very stable inside the layer with a sharp interface whereas the O contents is 46% in the center and looses its sharpness at the right interface. The Mg and O ratio is found to be 1.04 ± 0.13. The second interface seems to be more diffused. In both the cases we observed the presence of FeO and $Fe_2O$ more prominently at the second interface. If we look at the interface and measure the distance between the maximum and minimum positions of Fe content at both sides, as shown in Fig 5a (and 5b), the left interface has a thickness of 2nm (4nm) whereas the right of 7 nm (9nm). So we can conclude that the second interface is always oxidised and has a higher inter diffusion than the first one. One noticeable thing is that in the thin and thick layer the FWHM of $Fe_2O$ is about 2.5 nm. This is probably due to the almost same laser contribution during the two analyses.

3.2. Analysis with Green laser ($\lambda_1$=515 nm)

In the previous section, using UV laser we observe an oxidation of the second metal-insulator interface whatever the thickness of the oxide layer. To check whether this oxidation is related to the analysis condition as the laser wavelength, the two MgO samples were analysed using green laser.

Using the same protocol for UV the thin MgO layer is analysed under green wavelength. The base tip temperature was kept at 80K. The evaporation rate, 0.005-0.003 atoms/pulse and the laser fluence of 2.4 J/m$^2$ are kept constant through out the experiment. Fig 6a shows the reconstruction of atom distribution of the thin layer. Here also we can see the thickness of MgO correctly reconstructed as in Fig 4a. The layer is curved in the whole volume so for the composition profile calculation a square selection is done with a parallelepiped volume of 8×8×40 nm$^3$ having a base parallel to the layer. Fig 6b shows the 3D mapping of the thick layer. Again, a parallelepiped with a volume of 8×8×50 nm$^3$ is taken for the concentration profile calculation. Similar oxidation and depleted zone as observed at the second interface, as already reported in the previous section. Fig 7 shows the two concentration profiles of MgO layer of two different thicknesses.

For the thin layer, Fig 7a, shows the inter-diffusion and inter-mixing of first and second interfaces with MgO layer and $Fe_xO_y$ layer. Because of the depletion zone at the second interface, the Fe content is found to be less in the second layer as compared to the first layer. The oxygen content is found quite less than Mg. The ratio of Mg and O is found 1.14±0.21. However for thick layer (Fig 7b), the ratio between Mg and O was found to be 1.08± 0.12 which is also comparable with UV analysis. The depth of left interface is 2.5 nm (3nm) whereas at the right it is 8 nm (8nm) for the thin (thick) layer. Hence, as already observed using UV laser the second interface is always oxidised and diffused more than the first one.

As a conclusion the oxidation is always present whatever might be the laser wavelength as well as the thickness of the layer. The depth of the oxidation on both sides is more or less the same for all the cases. Moreover the FWHM of $Fe_2O$ is about 2.5 nm in both thin and thick layer using UV or green laser.

Therefore, the cause of oxidation was thought to be related to sample preparation. However, to clear this uncertainty, we performed another analysis using green laser on the thin layer by changing the laser fluence to a much higher value (6.3 J/m$^2$) and a comparatively higher evaporation rate and had a look at the concentration profile.

In Fig 8, we can see the same features of concentration as observed for earlier cases and the same depth for the interfaces. Moreover, in this case, the FeO peak is almost negligible and the concentration of $Fe_2O$ is very high. Here the FWHM of $Fe_2O$ is around 4 nm. Due to the different evaporation conditions (laser fluence, standing voltage), the ratio between the two oxides is changing already reported by Bachhav et al [26]. As a conclusion the thickness of the oxidation layer at the second interface (MgO/Fe) doesn't depend on the laser wavelength, laser fluence and the MgO thickness. Nevertheless we can not conclude that the oxidation is due to the sample preparation, because we should take into account the APT reconstruction aberration always present when two layers having different evaporation field are analysed. This is why we compared the analysis with numerical simulation of Fe/MgO/Fe evaporation behaviour.

3.3. Simulation and results

In an atom probe, the specimen is a sharply pointed needle with an end radius of R, is evaporated atom by atom while the impact positions of the projected ions on a detector are determined. The tomographic reconstruction is calculated using a simple back projection algorithm that takes into account the geometric features of the sample. This procedure, well adapted to homogeneous materials, founds its limitation when analysing nanostructure materials. The end form of the specimen changes continuously during the evaporation, depending on the structure of the analysed material, the experimental conditions and of the nature of elements present in the different phases. This can give rise to changes of the projection law between the detector and the specimen surface. Reconstruction aberrations are therefore often observed and can bias measurements [27]. In order to overcome this shortcoming, a modelling approach was developed [28, 29]. To understand the effects of the nanostructure on the back projection algorithm, the gradual evolution of the tip under field evaporation is modelled numerically. This evolution is obtained by calculating the electric field above the surface of the tip submitted to the DC voltage and by determining the induced sequence of evaporation. To obtain data corresponding to the result of an experimental analysis, the ion trajectories are calculated up to a virtual detector.

The field emitter is represented as a stack of atoms in the form of a cylinder terminated by a hemispherical cap of radius $R$ at the end of it. In order to obtain a completely filled space with no overlap between the atoms, the atomic volume is defined by the Wigner-Seitz cell of the chosen crystalline structure, which is a truncated octahedron for the body centred cubic structure. The electric field is introduced by a difference of electric potential $V$ between the tip and a counter electrode at a distance $D$ from the initial tip surface. From this, the potential $V(r)$ can be calculated in the entire simulated volume by solving numerically the Laplace equation:

$$\Delta V = 0 \tag{1}$$

From the potential distribution, the local electric field can be deduced in the entire volume. For each surface atom $n$ the electric field $F_n$ is measured at the surface of the corresponding Wigner Seitz cell. Field evaporation is the transition of a surface atom into an n-times charged ion under the influence of the external electric field $F$ ($F$ in the range 10-100 V/nm). This process is thermally activated process following Arrhenius law

$$K_n = \nu \exp\left(-\frac{Q_n(F)}{k_B T}\right) \tag{2}$$

where $Q_n$ is an activation barrier which depends on the field, $T$ the temperature and $k_B$ the Boltzmann constant. As a result, the lifetime $\tau_n$ of an atom at the tip surface submitted to a field $F_n$ is written as[30]:

$$\tau_n = \tau_0 \cdot e^{\frac{Q_n(F_n)}{k_B T}} \tag{3}$$

This equation resolves the evaporation sequence step by step, which in turn determine which atom to evaporate next. The probability for a surface atom to be chosen is inversely proportional to its lifetime $\tau$, which varies with the height of the activation barrier $Q_n$. The dependence of activation energy and the field $F_n$ is described in details in literature [27]. It differs from one atomic species to the other, but it is generally accepted that the energy barrier decreases monotonic with the electric field, and disappears for $F > F_e$, with $F_e$ the evaporation field of the chemical species concerned.

Even if no precise analytical expressions for the activation barrier of different elements $Q_n^{id}$ exists, a linear relationship of $Q_n$ with $F$ is generally accepted [27], so that it is possible to compare the lifetime of different chemical species assuming only the evaporation fields of the different elements. The evaporation fields for two layers are estimated from the voltage drop in the experiment (MgO has a evaporation field 20% less than Fe) .Taking into account all these assumptions, the lifetime of all surface atoms is calculated from the local electric field, normalized by the evaporation field of the layers identity. The chemical identity is known from the definition of the specimen, and the results of the simulation - mass, position, order of evaporation - can be compared to the experimental results. More details about the model can be found in previous publications [28, 29, 31]. To take into account the influence of the temperature simulation of the random influence of a thermally activated process at finite temperature was done for several values of temperatures. The final choice of the atom to evaporate next is performed by a quasi random process, where the evaporation probability is inversely proportional to the atom's life time. More details are reported in ref [32].

The FeMgOFe system was modelled by considering a tip composed of pure Fe with a layer of MgO perpendicular to the specimen axes. MgO was considered as a homogenous pure material with unique evaporation field. The crystalline structure was considered as bcc lattice. The evaporation field for Fe: 35 V/nm (1.0) and MgO: 28 V/nm (0.8) are taken. Fig 9 shows the simulation results which reveal a depleted zone (DZ) just after the MgO layer.

Fig 9(a) and Fig 9(b) show the 3D reconstruction at two different temperatures 200K and 400K respectively, and it is observed that the width of the depleted zone varies with temperature; it decreases of about 20% when the temperature doubles. This behaviour can be checked by experimental results obtained using two different laser energies. Fig 10 shows two experimental data with two different laser fluences corresponding to two different heating of the tip. The depleted zone (DZ) is estimated by the concentration profile of Fe and it is observed that the decrease of about 30% of depleted zone is possible with laser fluence almost double. Hence, the model can reproduce qualitatively the experimental results as it shows similar behaviour.

All the above results are a consequence of the discrepancies in the field evaporation of two different species, one having high field evaporation (Fe) and the other having low (MgO). So, when the evaporation goes from Fe to MgO, means from a higher field zone to a lower field zone, the evaporation process is smooth. But when the MgO layer is removed and the evaporation approaches the Fe layer, the evaporation process is almost stopped due to the higher field requirement. Because of this difference in the field, the tip shape changes a lot and the evaporation is not homogeneous anymore at the tip surface. The evaporation takes place only from the part at low field, and, from the rest, no evaporation until the required field is provided.

As a result a depleted area appears in the 3D construction just after the MgO layer but not before the MgO layer. These modifications in the tip structure can be simulated using above described model. Fig 11 depicts the change of the tip shape during field evaporation. Step 1 is the starting of the analysis which follows step 2 during evaporation process. At step 3 the MgO layer approaches and the radius of the tip is small. Finally for step 4 the radius is very big and the MgO is almost completely removed. As we can see at the step 4 the radius of the tip is very large compared to the radius of the tip at the beginning of the analysis and the surface is not protected by high field. As a result oxygen molecules from the vacuum rise from the shank and oxidise the surface and create several iron oxides which are identified in the mass spectra and 3D reconstruction. In the simulation results we can not predict these oxidation phenomena because, in our model only two chemical species were taken into account one having higher field and another having lower field evaporation. Any intermolecular interaction was neglected.

## 4. Conclusions

We have studied the metal-insulator-metal interfaces of Fe/MgO/Fe samples because the sharpness of these interfaces is strongly related to the TMR of the structure. We use APT to analyse two barriers having two different oxide thickness (MgO of 4 nm and 32 nm). The spatial resolution of APT is very high (less than 1 nm) hence the interfaces can be reconstructed with an atomic resolution. The APT analysis show the presence of an oxidation layer of the two interfaces however the second (MgO/Fe) interface has a relatively larger interdiffusion as compared to the first interface (Fe/MgO) with a larger oxidation depth. We demonstrate in this paper that this phenomenon is independent of laser wavelength, laser fluence and the thickness of the MgO layer. Considering the Fe and MgO as two atomic species having two different evaporation field (high evaporation field for Fe and 20% lower for MgO), we studied by numerical simulation of the field evaporation of Fe/MgO/Fe. Due to the two different evaporation fields, the tip shape changes during the evaporation process making the evaporation non-homogeneous on the tip surface and allowing the surface diffusion of O atoms and consequently the surface oxidation. It can be concluded that the interface oxidation is a consequence of field evaporation and can not be associated to the sample preparation method.

More efforts are in progress to upgrade our evaporation model to take into account intermolecular interaction in MgO layer. Furthermore we will be able to predict the iron oxidation and consequently we will be able to correct this APT aberration. Only in these condition APT will be correctly used to study interfaces having variable evaporation fields.

## Acknowledgments

We acknowledge the ESP Carnot, the TAPAS, the ANR and Cameca France for supporting our work. We would like to thank Ryota Gemma and Dr. Talat Al-Kassab from University of Gottingen for helping in MgO sample preparation.

Figures :

In the final printed version of paper.